\documentclass[twocolumn]{aastex631}

\usepackage{graphicx}
\usepackage{dcolumn}
\usepackage{bm}
\usepackage{hyperref}
\usepackage[mathlines]{lineno}
\usepackage{xspace}
\usepackage{multirow,dashrule}

\graphicspath{{./figs/}{./png/}}

\newcommand{\EQ}{\begin{equation}}
\newcommand{\EN}{\end{equation}}
\newcommand{\EQA}{\begin{eqnarray}}
\newcommand{\ENA}{\end{eqnarray}}

\newcommand{\Eq}[1]{Eq.~(\ref{#1})}

\newcommand{\App}[1]{Appendix~\ref{#1}}

\newcommand{\FFig}[1]{Figure~\ref{#1}}

\newcommand{\Tab}[1]{Table~\ref{#1}}

\newcommand{\mean}[1]{\overline#1}

{}
{}
{}

{}
{}
{}
{}
{}
{}
{}
{}
{}
\newcommand{\meanBB}{\overline{\mbox{\boldmath $B$}}{}}{}
{}
{}
{}
{}
{}
{}
{}
{}
\newcommand{\meanUU}{\overline{\bm{U}}}

{}
{}
{}

\newcommand{\meanB}{\overline{B}}

\newcommand{\meanU}{\overline{U}}

\newcommand{\flucBB}{\mbox{\boldmath $B^\prime$}{}}{}
{}
\newcommand{\flucuu}{\mbox{\boldmath $u^\prime$}{}}{}
{}
{}

{}

{}
{}

\newcommand{\MR}{{GCD}\xspace}
\newcommand{\PGCD}{P_{\rm cyc}^{\rm \MR}}
\newcommand{\uu}{\mbox{\boldmath $u$} {}}

\newcommand{\nab}{\mbox{\boldmath $\nabla$} {}}

\newcommand{\ddelta}{\mbox{\boldmath $\delta$} {}}
\newcommand{\ggamma}{\mbox{\boldmath $\gamma$} {}}
\newcommand{\aalpha}{\mbox{\boldmath $\alpha$} {}}
\newcommand{\bbeta}{\mbox{\boldmath $\beta$} {}}
\newcommand{\kkappa}{\mbox{\boldmath $\kappa$} {}}

\newcommand{\EMF}{\mbox{\boldmath ${\cal E}$} {}}

\def\urms{u_{\rm rms}}

\newcommand{\W}{\,{\rm W}}
\newcommand{\V}{\,{\rm V}}

\newcommand{\T}{\,{\rm T}}
\newcommand{\G}{\,{\rm G}}

\newcommand{\K}{\,{\rm K}}
\newcommand{\g}{\,{\rm g}}
\newcommand{\s}{\,{\rm s}}

\newcommand{\m}{\,{\rm m}}

\newcommand{\J}{\,{\rm J}}

\newcommand{\A}{\,{\rm A}}

\shorttitle{Investigating dynamos with mean-field models}

\shortauthors{Warnecke et al.}

\begin{document}

\title{Investigating global convective dynamos with mean-field models:\\ full spectrum of turbulent effects required}

\author[0000-0002-9292-4600]{J\"orn Warnecke}
 \affiliation{Max-Planck-Institut f\"ur Sonnensystemforschung,
Justus-von-Liebig-Weg 3, D-37077 G\"ottingen, Germany; warnecke@mps.mpg.de
}%

\author[0000-0001-9840-5986]{Matthias Rheinhardt}
\affiliation{
Department of Computer Science, Aalto University,
PO Box 15400, FI-00\ 076 Espoo, Finland
}%

\author[0000-0003-3317-5889]{Mariangela Viviani}
\affiliation{%
Department of Physics, University of Calabria,
I-87036, Rende (CS), Italy
}%
\affiliation{%
Max-Planck-Institut f\"ur Sonnensystemforschung,
Justus-von-Liebig-Weg 3, D-37077 G\"ottingen, Germany
}%

\author[0000-0002-1331-2260]{Frederick A. Gent}
\affiliation{
Department of Computer Science, Aalto University,
PO Box 15400, FI-00\ 076 Espoo, Finland
}%
\affiliation{School of Mathematics, Statistics and Physics,
      Newcastle University, Newcatle upon Tyne NE1 7RU, UK}

\author[0000-0003-3317-6777]{Simo Tuomisto}
\affiliation{
Department of Computer Science, Aalto University,
PO Box 15400, FI-00\ 076 Espoo, Finland
}%

\author[0000-0002-9614-2200]{Maarit J. K\"apyl\"a}
\affiliation{
Department of Computer Science, Aalto University,
PO Box 15400, FI-00\ 076 Espoo, Finland
}%
\affiliation{%
Max-Planck-Institut f\"ur Sonnensystemforschung,
Justus-von-Liebig-Weg 3, D-37077 G\"ottingen, Germany
}%
\affiliation{%
Nordita, KTH Royal Institute of Technology \& Stockholm University,
Hannes Alfv\'ens v\"ag 12, SE-11419 Stockholm, Sweden
}%

\begin{abstract}
The role of turbulent effects for dynamos in the Sun and stars
continues to be debated.
Mean-field (MF) theory provides a broadly used framework to connect
these effects to fundamental magnetohydrodynamics.
While inaccessible observationally, turbulent effects can be directly
studied using global convective dynamo (\MR) simulations.
We measure the turbulent effects in terms of turbulent transport
coefficients, based on the MF framework,
from an exemplary \MR simulation using the test-field method.
These coefficients are then used as an input into an MF model.
We find a good agreement between the MF and \MR solutions,
which validates our theoretical approach.
This agreement requires all turbulent effects to be included, even those 
which have been regarded as unimportant so far. 
Our results suggest that simple dynamo models,
 as are commonly used in the solar and stellar community,
 relying on very few, precisely fine-tuned turbulent effects,
may not be representative of the full dynamics of
dynamos in global convective simulations and astronomical objects.
\end{abstract}


\section{Introduction}

The magnetic fields of the Sun and other cool stars are generated by a
dynamo mechanism operating in their interiors.
Despite plentiful observations, over an extensive history with many at
high resolution, the nature of the solar dynamo is not yet fully understood.
One of the difficulties lies in the 
poor knowledge of the turbulent effects,
which are expected to play an important part in the magnetic field
generation.
These effects are often described by a parameterization based on mean-field (MF) theory.
Because of their intricacy, however, 
it is common in the solar context to simplify the parameterizations
such that they can be fine-tuned to fit some of the magnetic field
observations \citep[e.g.][]{KJMCC14,CS15}.
As an alternative approach, global convective dynamo (\MR)
models can be used to self-consistently generate these turbulent effects.
While these models have parameters 
far from real astrophysical objects, they currently represent the best laboratories to this end.
Nevertheless, investigating the nature of dynamos in \MR is very
challenging and needs an analysis tool, which connects properties of
the \MR to established dynamo theories. 
In the recent years the test-field method (TFM) has become 
such a well-established tool to measure the turbulent transport
coefficients (TTCs),
quantifying the turbulent effects, in such \MR models
\citep{SRSRC05,SRSRC07, SPD11,SPD12,Schr11, W18, WRTKKB17, VKWKR19,WM20}.
Already in these studies, it was found that the turbulent effects
play an important part in the magnetic field evolution.
However, to show that the TTCs measured by the TFM
capture the most important details of the magnetic field evolution, the
coefficients need to be employed in an MF
model and the results to be compared with the {\MR} simulations.
Furthermore, only with the use of an MF model one will be able to
pinpoint which of these turbulent effects are essential 
for a full understanding of dynamo action in the {\MR}.
Steps in this direction, beyond the works of Schrinner et al., have been performed using a
small subset of TTCs
($\alpha$ tensor and turbulent pumping)
obtained from a \MR simulation using the 
singular value decomposition (SVD)
method \citep[][]{SCB13,SC20}.

In our work, we use in an MF model the TTCs of an exemplary {\MR}
model, the magnetic field evolution of which shows similarities to the
Sun, to investigate whether or not this evolution can be
reproduced. Furthermore, we will investigate which minimal subset of the coefficients is
essential to reproduce the dynamo solution and hence determine whether or not
it can be described by any of the commonly employed simple dynamo
models. We will conclude by discussing the further implications of the
results.

\section{Models and Methods}
\label{sec:model}

\begin{figure*}[t!]
\begin{center}
\includegraphics[width=1.9\columnwidth]{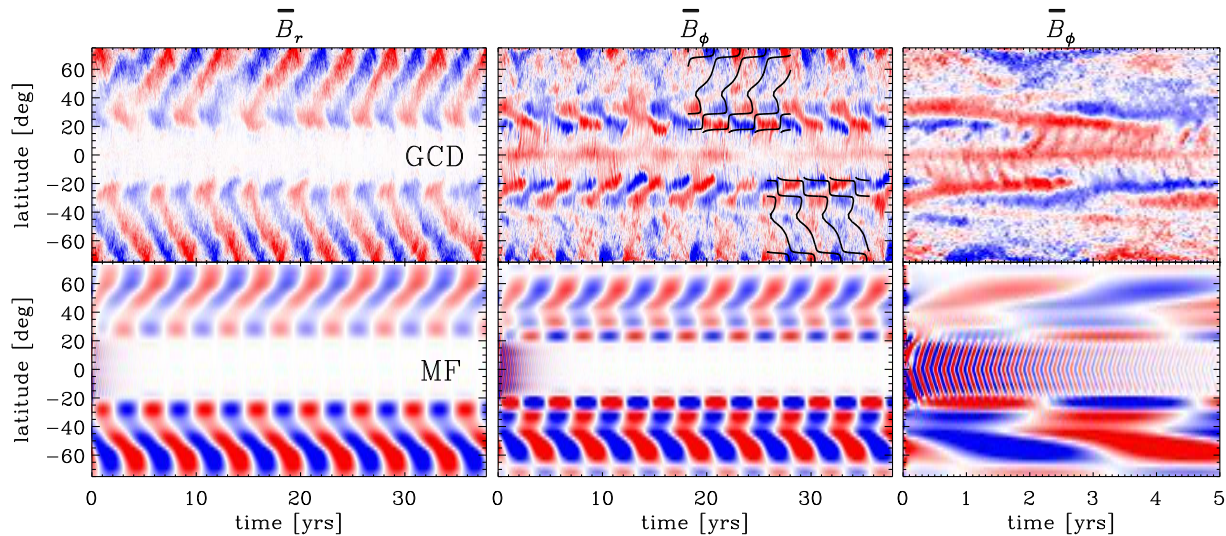}
\end{center}\caption{
Time--latitude (butterfly) diagrams of mean radial, $\meanB_r$, and
azimuthal, $\meanB_\phi$, magnetic field from \MR (top) and the MF
model (bottom) at fractional radius 0.95.
The rightmost
panels show a zoom-in to the first five years of the middle panels.
The TTCs are symmetrized,
and $\aalpha$ is scaled by 1.5, 
while exponential growth has been compensated for clarity.
The color range is cut to make the 
northern hemisphere more visible.
Black 
lines: zero contours of $\meanB_\phi$
from the MF model at the same time.
See \App{butAS} for butterfly diagrams of the
corresponding pure-parity solutions.
}\label{but}
\end{figure*}

We analyze Run M5 of \cite{W18} and \cite{WM20}, a {\MR} simulation
in a spherical shell.
It has a rotation rate roughly four times higher 
 than the Sun, in terms of the Coriolis number\footnote{The Coriolis number is defined as 
$\Omega\Delta R/\pi\urms $ 
with the overall angular frequency $\Omega$,
the volume-integrated rms velocity $\urms$ and the thickness of the convective shell $\Delta R$.
For the Coriolis number of the Sun, see, e.g. \cite{SB1999}.
}
and a Rayleigh number%
{\footnote{The Rayleigh number is defined based on the mean entropy
  gradient in the middle of the convection zone determined from the
  hydrostatic counterpart of the run; see \cite{KMCWB13} for details.}}
two orders of magnitude larger than the critical one for
the onset of convection \citep{WRTKKB17}; see also \cite{KMCWB13} and \cite{W18} for
more details on this run and its comparison with solar parameters.
The axisymmetric (azimuthally averaged)
part of the generated magnetic field shows
rather regular oscillations with a magnetic cycle period 
$P_{\rm cyc} = \PGCD \equiv 4.4\pm0.6$ yr \citep{W18}, 
 and exhibits both
equatorward and  poleward branches of field
migration in the butterfly (time--latitude) diagram, 
thus capturing main solar cycle
features; see \FFig{but}.
We note here that this is not a one-to-one representation of the Sun and its dynamo, but a good example of state-of-the-art models of
  solar-like stars in the scientific community \citep[e.g.][]{ABMT15,SBCBN17}.
A second, weaker dynamo mode with a
much shorter period of about $0.11$ yr is present
at low latitudes near the surface; see \FFig{but} and
\cite{KKOBWKP16} for a detailed discussion.

Our MF approach employs azimuthal averaging,
indicated by an
overbar, which adheres to the Reynolds rules;
fluctuating fields are indicated by primes.
The MF induction equation reads
\begin{equation}
\partial_t \meanBB = \nab\times \left(\meanUU \times \meanBB + \EMF\right) - \nab\times \eta\nab\times\meanBB,
\label{eq:meanInduc}
\end{equation}
where $\meanUU$ and $\meanBB$ are mean flow and mean magnetic field,
respectively, and $\eta$ is the magnetic diffusivity.
To establish the MF model, a parameterization of the mean
electromotive force $\EMF=\overline{\flucuu\times\flucBB}$
in terms of the mean field itself is crucial.
Employing Taylor expansion, leaving out time derivatives and
restricting to first-order spatial derivatives, a commonly quoted
ansatz reads \citep{KR80}
\begin{equation}
\EMF=\aalpha\cdot\meanBB+\ggamma\times\meanBB 
-\bbeta\cdot\left(\nab\times\meanBB\right)
-\ddelta\times\left(\nab\times\meanBB\right) 
-\kkappa \cdot\left(\nab\meanBB\right)^{(s)},
\label{EMF}
\end{equation}
where $(\nab\meanBB)^{(s)}$ is the symmetric part of the (covariant)
derivative tensor of $\meanBB$.
The most general representation of $\EMF$ at some position $(r,\theta)$
and time $t$ would involve a convolution integral over a neighborhood of 
$(r,\theta,t)$, thus covering nonlocal and memory effects \citep{KR80}.
In contrast, \Eq{EMF} is completely instantaneous in time and only rudimentarily 
nonlocal in space.

$\aalpha$ and $\bbeta$ are symmetric rank-two tensors, $\ggamma$ and
$\ddelta$ are vectors, and $\kkappa$ is a rank-three tensor, with
symmetry $\kappa_{ijk}=\kappa_{ikj}$. These five tensors 
can be associated with
different turbulent effects important for the magnetic field
evolution: the $\alpha$ effect \citep{SKR66}
can lead to field amplification via helical flows, e.g. in stratified rotating convection,
the $\gamma$ effect describes turbulent pumping of the mean magnetic
field, in analogy to a mean flow.
 $\bbeta$ describes turbulent diffusion; and the $\delta$, or
R\"adler effect \citep{KHR69}, can lead to dynamo action in the
presence of, e.g., $\alpha$ effect or shear, but not alone \citep{BS05}. 
The physical interpretation of $\kkappa$ is yet unclear, 
but quite generally it may contribute to both amplification and diffusion of
$\meanBB$.
More details on these effects and its parameterization can be found, e.g., in \cite{KR80} and \cite{BS05}.

Accounting for all symmetries in \Eq{EMF} and $\partial_\phi \meanB_{r,\theta,\phi}=0$,
a total of 27 coefficients must be identified to close \Eq{eq:meanInduc}.
Note that due to the axisymmetry of $\meanBB$, the representation of
\Eq{EMF}  is nonunique. 
In particular, components of $\kkappa$
can be recast into components of $\bbeta$. Here we have chosen a
formulation of $\bbeta$ and $\kkappa$ that maximizes the number of
vanishing entries in $\kkappa$, that is, allocates as much information
on the diffusive aspects of turbulent transport as possible in
$\bbeta$, thus facilitating physical interpretation; see
\cite{VKWKR19} for details.

To determine the required 27 coefficients we apply the 
quasi-kinematic
TFM to the original
\MR model.
The original TTCs are already published \citep{W18,WM20} and available
online.\footnote{\url{http://doi.org/10.5281/zenodo.3629665} \citep[see also][for TTCs of a very similar run]{WRTKKB17}}
The TFM, as utilized in \cite{SRSRC05,SRSRC07} and \cite{WRTKKB17}, 
requires nine additional realizations of the induction equation for the
fluctuating magnetic field $\flucBB$ to be solved simultaneously with the \MR
simulation.
They are obtained by replacing $\meanBB$ in $\nab \times\big(\flucuu\times\meanBB\big)$ by one out of nine
linearly independent
{\it test fields} $\meanBB^{(i)}$, $i=1,\ldots,9$,
while the velocity $\uu=\meanUU+\flucuu$ is taken directly from the \MR simulation
\citep[see][for details]{SRSRC07}.
Employing the corresponding electromotive forces
$\EMF^{(i)} = \overline{\flucuu\times\flucBB^{(i)}}$, 
a uniquely solvable linear equation system
for the coefficients in \Eq{EMF} can be formed.

In the stationary, saturated state of the \MR simulation
the magnetic field
acquires dynamically significant strength.
Hence, the velocity and TTCs derived from it are already
magnetically quenched, that is, they differ from their counterparts in the
nonmagnetic (or kinematic) state.
Consequently, the MF model of \Eq{eq:meanInduc}
is, strictly speaking, valid only for the mean field that is observed in the
\MR simulation \citep{BCDHKR10}.

We subject the TTCs to the following preprocessing steps:
firstly, we exclude the noise arising from large variations at small time scales
\citep{WRTKKB17} by averaging over their full time series;
secondly, to damp spatial fluctuations,
we apply a Gaussian smoothing in $\theta$
using a kernel with a standard deviation
equal to the grid spacing;
thirdly, we remove negative values from the diagonal $\bbeta$ components and
apply a lower threshold of $-0.7\eta$ for
$\kappa_{\phi r\theta}$ to avoid unphysical local instabilities.
For consistency, $\meanUU$ is time averaged, too.
Additionally, we  reduce the original resolution of the coefficients from
$180\times256$ (radial $\times$ latitudinal) to $40\times64$ for computational efficiency.

The set of equations for \MR
admits solutions with ``pure" equatorial symmetries, that is, equatorially symmetric 
velocity, density, and entropy fields, combined with an either
symmetric (S) or antisymmetric (A) magnetic field.
The degree of equatorial symmetry of a field is quantified by the {\it parity}
$P$, which takes values between $-1$ for A and $+1$ for S.
In the \MR simulation, the parity of the magnetic field
continuously varies between $+1$ and $-1$, indicating that 
A and S dynamo modes of similar
strength are competing for dominance \citep{KKOBWKP16}; see also \App{butAS}.
Each of the measured TTC components, however,
has near pure parity of one or the other \citep{WRTKKB17}. 
Therefore, to study the competing pure modes in isolation in the MF model we 
employ TTCs,
properly symmetrized so as to restrict their parity to,
respectively, $P=\pm1$.\footnote{A 
scalar $F$ is said to be symmetrized for parity 1 by
$(F(r,\theta)+F(r,\pi-\theta))/2$ and for 
parity $-1$ by
$(F(r,\theta)-F(r,\pi-\theta))/2$, while for a vector field the $r$ and $\phi$
components have  the same parity as the field as a whole,  
but the $\theta$ component has opposite parity.}

For the MF simulations we solve \Eq{eq:meanInduc} using the 
preprocessed $\meanUU$ and TTCs and
the same $\eta$ as in the \MR model without employing any quenching terms.
Hence, the MF model is entirely linear, and by virtue of the temporal constancy
of all its coefficients, all solutions show exponential behavior, in general with a complex increment.
The mean magnetic field is initialized by a weak random seed.

All simulations were performed using the {\sc Pencil Code}
\citep{PC21}; see \cite{SimoMT} for details of its MF module.

\section{Results}

First we look at an MF solution, 
the parity of which is not restricted while the
symmetrized TTCs are employed, and compare with the one of the \MR
model.
If the $\alpha$ tensor is scaled by a factor $f_\alpha$ between 
$1.40$ and $1.525$ we find growing oscillatory solutions,
which resemble the \MR solution very well, as shown in
\FFig{but}. Their periods of $4.5$--$4.9$ yr are in close agreement with
$\PGCD\equiv 4.4\pm0.6$ yr \citep{W18}.
The butterfly diagram is well reproduced, too:
the poleward migrating $\meanB_r$
pattern has the same shape and slope as in the \MR
simulation. 
In $\meanB_\phi$, the agreement of the pattern shapes is also striking as
signified by the zero contours of the MF solution plotted over the 
\MR one. Deviations are visible at the highest latitudes where the test-field
measurements are likely to be contaminated by the unphysical
latitudinal boundaries.
Also, at midlatitudes patches of equal polarity appear not to be
equatorwards connected as in the \MR solution.

Like the \MR solution, this MF one is neither purely antisymmetric nor
symmetric about the equator.
See \App{butAS} for details of the corresponding pure-parity solutions of the
MF model.
The zoom-in to the early phases of the MF model (see the lower rightmost
panel of \FFig{but}) shows the high-frequency, poleward 
migrating mode
with roughly $0.11$\,yr period length
at low latitudes.  
Field migration and period of this
mode match closely those of
the \MR.
While it is regular in the MF model,
it appears temporally incoherent
in \MR, agreeing with the findings of \cite{KKOBWKP16} for a very similar run. 
Given the absence of nonlinearities, this mode becomes subdominant in the
MF model.
We note here that the absence of nonlinearities also implies that there is
no interaction between the two pure-parity modes of the MF model.

\begin{figure}[t!]
\begin{center}
\includegraphics[width=\columnwidth]{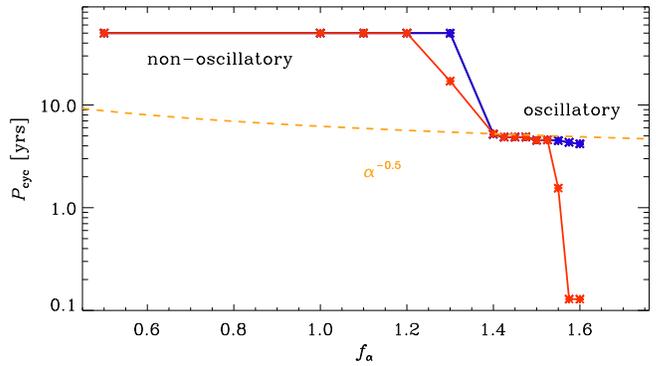}
\end{center}\caption{
Cycle period $P_{\rm cyc}$ as function of the scaling
factor $f_\alpha$.
Red: S, blue: A solutions.
Nonoscillatory solutions have been assigned $P_{\rm cyc}=50$ yr.
Dashed line: 
scaling of $P_{\rm cyc}\sim \alpha^{-0.5}$ according to
a Parker dynamo wave.
}\label{alp_sc}
\end{figure}

\begin{figure*}[t!]
\begin{center}
\includegraphics[width=\textwidth]{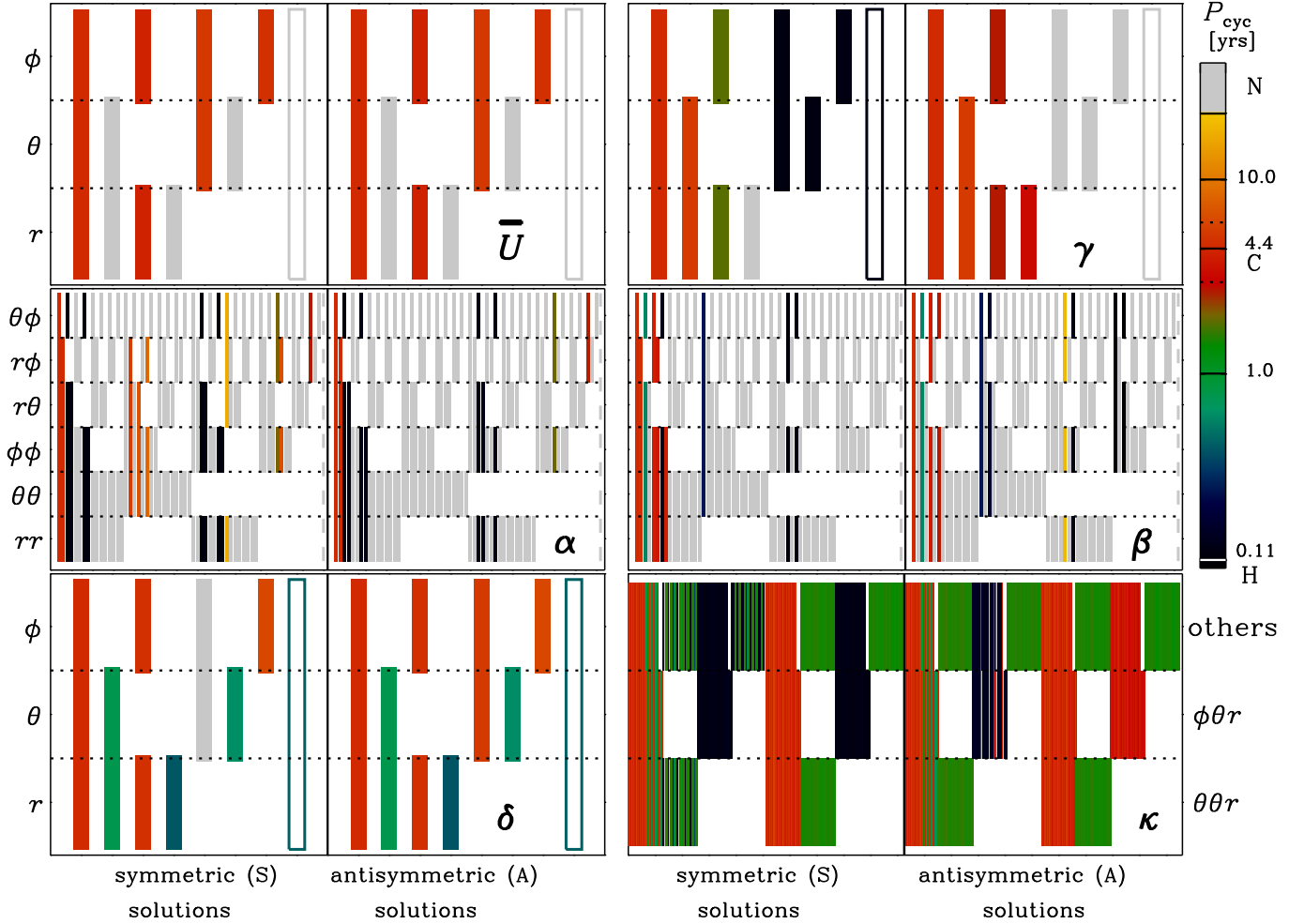}
\end{center}\caption{
S and A solutions for all combinations of turned-on components for each coefficient 
tensor with  $P_{\rm cyc}$ coded by colors.
Gray: nonoscillatory (N), red: correct oscillatory (C), and dark blue: high-frequency oscillatory (H) solutions.
For $\kkappa$, ``others'' includes all coefficients other than $\kappa_{\phi\theta r}$ and $\kappa_{\theta\theta r}$.
Empty boxes or dashed vertical lines: all components are turned off
(not shown for $\kkappa$).
Dotted lines in color bar: 
interval (3.0\,\ldots\,6.0) yr.
}\label{coeff}
\end{figure*}

We have repeated some of the MF runs with the nonsymmetrized TTCs.
Their slight hemispheric asymmetries are sufficient to excite a strongly
asymmetric eigenmode, such that one hemisphere exhibits a magnetic field
up to two orders of magnitude stronger than the other. 
The \MR simulation does not show such strong 
disparities, which we attribute to the nonlinearities in the \MR\ providing
a self-regulation mechanism: whenever significant disparity occurs, the hemisphere showing the higher
dynamo efficiency would also experience a stronger back-reaction of the magnetic
field on the flow (quenching).
Thus, dominance of one hemisphere likely cannot persist for long.
Instead, both the TTCs and the mean field will stay close to a
state of nearly pure parity.
The MF model, being linear, cannot provide such self-regulation.
Applying symmetrized coefficients, however, is sufficient to maintain 
consistent growth between the hemispheres.

The consistencies between the MF and \MR solutions prove that in our
case $\EMF$, i.e. the contributing turbulent effects, are well
described by the parameterization of \Eq{EMF}, meaning that higher-order terms, and
also scale dependence and memory effect, can be neglected.
Another surprising aspect is that employing only the time-averaged,
smoothed, and downsampled TTCs,
that is, ignoring their large temporal and small-scale spatial variations, 
is not detrimental to the agreement with the \MR solution,
but might explain the necessity of a moderate
upscaling of $\aalpha$.  

For values of $f_\alpha<1.40$,
the solutions are growing with a dominantly
nonoscillatory field, see \FFig{alp_sc}.
Interestingly, the oscillation period is closest to that of
the \MR model,
when the growth rates of the A and S solutions are
nearly identical,
hence enabling a mixed solution.
For $f_\alpha > 1.525$,
the oscillation period of the S mode is strongly reduced, closely matching 
that of  the high-frequency mode reported for the \MR model. 
Oscillation periods close to $\PGCD$
depend only weakly on $f_\alpha$, consistent with the expected scaling of the period of an
$\alpha\Omega$ dynamo wave, $P_{\rm cyc}\sim\alpha^{-0.5}$ \citep{P55,Yos75}; 
see  \FFig{alp_sc}.
This agrees with earlier findings that
direction and period of dynamo waves in \MR
models of this kind can be well explained by the Parker--Yoshimura rule \citep{WKKB14,WRTKKB17,W18}.

Next we analyze which of the TTCs and
mean flow components are essential for reproducing
the \MR solution.
For this we perform around 2000 MF simulations with $f_\alpha=1.5$,
where we turn on/off the various components of a certain TTC tensor
at a time, while fixing all the others at their nominal values,
to investigate the changes in the resulting MF solution.
To classify the solutions, we define three different classes based on their period; see \FFig{coeff}. The first
class (C for ``correct") has a period in the interval (3.0\,\ldots\,6.0) yr around 
$\PGCD$, the second (N) has dominantly nonoscillatory solutions, 
the third (H for ``high frequency") has oscillatory solutions, close to
the near-surface high-frequency dynamo mode with $P_{\rm cyc}=0.11$ yr.
For a set of coefﬁcients to be
considered essential to reproduce the \MR solution
we require that both the S and A solutions fall into class C (details in
\Tab{ncyl} of \App{TabC}).

First, we consider the effect of $\meanUU$.
As illustrated in \FFig{coeff}, only solutions with a nonzero
$\meanU_\phi$ are oscillatory with a period close to  $\PGCD$;
all other cases have only nonoscillatory
solutions. This means
that the meridional circulation ($\meanU_r$,$\meanU_\theta$) has a negligible contribution,
while differential rotation
($\Omega$ effect) is crucial, in clear contrast to the often invoked
flux-transport dynamo models \citep[e.g.,][]{KJMCC14}, 
where the meridional circulation plays an essential part.
All solutions with the full $\alpha$ tensor (or at best missing
$\alpha_{\theta\phi}$) fall into class C.
Most of the other solutions fall either in the N or H class; see \FFig{coeff}.
Given the prime importance of  $\aalpha$
and $\meanU_\phi$, we conclude
that an $\alpha^2\Omega$-type dynamo is operating in our
model.
This is in agreement with the analysis performed in \cite{WM20} for this
run, where 
the $\alpha$ and $\Omega$ effects
were estimated to be
comparable in generating the toroidal field; see their Fig. 13.
We further point out that the high-frequency mode (class H) is already
present if from all the TTCs only certain $\aalpha$ components are
active, and therefore we identify it as an $\alpha^2$ dynamo mode in
agreement with earlier findings \citep{WRTKKB17}.

Next, focussing on  $\ggamma$ and $\ddelta$, 
we find that for obtaining  
class-C solutions,
$\gamma_r$ and $\gamma_\theta$ are important
as without one of them only class-N or class-H
solutions (additionally one with $P_{\rm cyc} \simeq1.8$ yr) arise.
Likewise, class-C solutions require $\delta_r$ and $\delta_\phi$;
otherwise, only class-N solutions, or periods between classes C and H,
are possible; see \FFig{coeff}.

The plentiful spectrum of solutions obtained by varying the selection of $\bbeta$
components (\FFig{coeff}) includes a large fraction that is
nonphysically unstable\footnote{This refers to extremely localized
  rapidly growing field structures on the grid scale, typically
  appearing as a checkerboard pattern and being characteristic for
  negative diffusivity.}.
This is because arbitrarily dropping components of $\bbeta$ can in general destroy its
positive definiteness.
Interestingly, only when at least the diagonal components plus $\beta_{r\phi}$ and $\beta_{\theta\phi}$
are active, a class-C solution 
can be reproduced.
However, $P_{\rm cyc}=\PGCD$ 
requires the full $\beta$ tensor.  

From approximately
 1500 solutions with various selections of
$\kkappa$ components, 
we find that around a third fall into class C, requiring
the components $\kappa_{\theta\theta r}$ and
$\kappa_{\phi\theta r}$ to be turned on; see \FFig{coeff}.
Other solutions populate
the H class or show
periods of $P_{\rm cyc}\simeq1.7$\,yr.
Interestingly, many solutions have growth rates larger than that of the full MF model.
Thus the $\kappa$ tensor provides additional
diffusion to the system.
Despite the fact that $\kkappa$ is often discarded,
we thus find that at least two components are
essential for reproducing the \MR solution.

To conclude, we find that a minimum set 
of  TTCs, capable of reproducing the \MR solution,
requires $\meanU_\phi\,(\Omega)$,
the full $\aalpha$ (except $\alpha_{\theta\phi}$) and $\bbeta$ tensors,
$\gamma_r$ and $\gamma_\theta$, 
$\delta_r$ and $\delta_\phi$, and $\kappa_{\theta\theta r}$ and 
$\kappa_{\phi\theta r}$.
For further testing this, we performed an MF run with the minimal
set of coefficients and found a very good match with the \MR simulation;
see \Tab{ncyl} and \App{butmin}. 

\section{Conclusion}

In our work, we find that the full spectrum of turbulent effects
($\Omega$, $\aalpha$, $\bbeta$, $\gamma_r$, $\gamma_\theta$,
$\delta_r$, $\delta_\phi$, $\kappa_{\theta\theta r}$,
$\kappa_{\phi\theta r}$) is required to reproduce the \MR solution.
This has two noteworthy implications.

First, all our findings agree
with previous works \citep{WKKB14, WRTKKB17, W18, WM20}
insofar as we found that the oscillation
period is mostly controlled by the $\alpha$ and $\Omega$
effects and not by the advection due to
meridional circulation.
This is in stark contrast to simplified flux-transport dynamo
  models \citep[e.g.][]{KJMCC14},
which are commonly invoked to describe the solar dynamo and
  rely crucially on meridional flow.
Furthermore,  our results support the concepts of turbulent pumping
($\ggamma$) and  R\"adler effect ($\ddelta$) to be essential
for the dynamo
\citep[e.g.][]{SRSRC05,SPD12,SB15b,Shi16,GKW17,WRTKKB17,VKWKR19,GE20,WM20,P21}.

Second, and more importantly, our work shows that 
most probably, all turbulent effects considered here
are required in the generation of the magnetic field,
and therefore are all needed to understand the
dynamo operating in \MR simulations
\citep[in agreement with][]{SRSRC05}.
Our results hence suggest that simple models based on a 
handful of fine-tuned coefficients, as commonly used in the solar and stellar 
context, miss crucial effects at play in \MR models.

Our results are in agreement with the study of \cite{SCB13}, where
the authors found that the contributions of the full $\aalpha$ and
$\ggamma$ are important to reproduce their \MR solution. However,
they did not consider the effect of $\bbeta$, $\ddelta$
and $\kkappa$.
In addition, their TTCs have been measured using the SVD method, which
has been shown to give misleading results if applied to our \MR
simulations \citep{WRTKKB17}.

Given the severely limited observability of stellar interiors, even
yet in the case of the Sun, such models are currently the only
laboratories for quantifying the turbulent effects. 
Furthermore, simply investigating the electromotive force \citep[e.g.][]{ABMT15,SBCBN17} or
measuring the TTCs and evaluating the
relative strengths of the corresponding effects is insufficient.
Only by examining these coefficients
within an MF model, one can
comprehend the dynamo at work in the \MR.  
Even though the \MR simulations do not attain realistic parameters
yet, the understanding of their dynamos is a fundamental 
step (together with observations and experiments)
toward understanding dynamos in many astrophysical objects, in particular in
the Sun and other stars.

\begin{acknowledgments}
We thank the reviewer for constructive comments on the manuscript.
Simulations have been conducted on supercomputers at
GWDG, the Max Planck supercomputer at RZG in Garching,
 and facilities hosted by the CSC--IT
Center for Science in Espoo, Finland.
J.W.\ acknowledges funding by the Max-Planck/Princeton Center for
Plasma Physics.
M.V.\ acknowledges support from the HPC-EUROPA3 project (INFRAIA- 2016-1-730897),
supported by the EC Research Innovation Action under the H2020 Programme.
This work has received funding from the European Research Council
(ERC) under the European Union's Horizon 2020 research and innovation
program (grant agreement No. 818665 ``UniSDyn''), and has been
supported from the Academy of Finland Centre of Excellence ReSoLVE
(project number 307411).
\end{acknowledgments}

\bibliographystyle{aasjournal}
\bibliography{paper}


\appendix
\section{Time--Latitude Diagram of Pure A and S Solutions}
\label{butAS}

\begin{figure*}[h!]
\begin{center}
\includegraphics[width=0.7\textwidth]{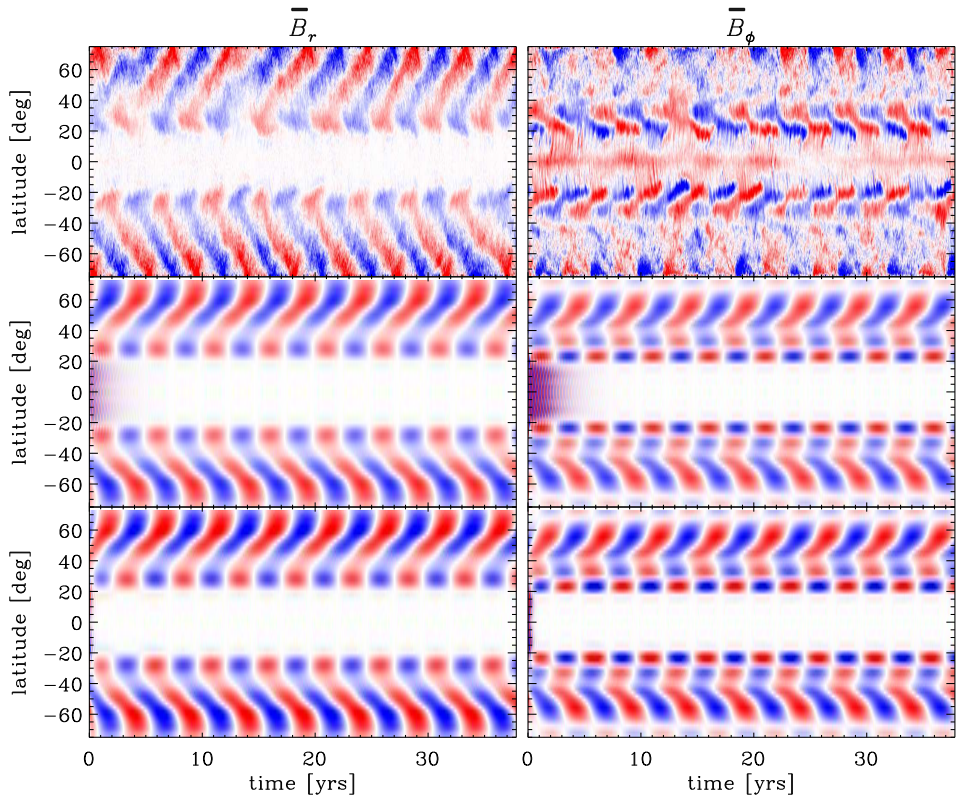} 
\end{center}\caption{
Time--latitude (butterfly) diagrams of mean radial, $\meanB_r$, and
azimuthal, $\meanB_\phi$, magnetic field from the \MR (top) 
together with the symmetric (S) and  antisymmetric (A) solutions 
of the MF model (middle and bottom)
 at fractional radius 0.95 (similar to \FFig{but}).
The TTCs are symmetrized,
and $\aalpha$ is scaled by 1.5, 
while exponential growth has been compensated for clarity.
}\label{butS1}
\end{figure*}

We show in \FFig{butS1} the purely symmetric (S) and antisymmetric (A) solutions of the MF model together with the \MR simulation.
The solution shown in \FFig{but} should be regarded as a weighted sum of these two solutions.
As the MF model is linear, the weight depends on the initial condition, and is, in this sense, arbitrary.
In that particular MF run, the parity, computed over all depths,
 is $-0.6$, indicating a larger contribution from the A than the S solution.
This also closely matches the \MR simulation, where the parity switches back and
forth between $-1$ and $+1$ with a period of around 20 yr;
the average parity is $-0.31\pm0.3$ \citep[see also][for a detailed analysis of the parity for a
  similar \MR simulation]{KKOBWKP16}.
We note that, while the dominating modes of the A and S solutions have equal growth rates, the high-frequency mode has a clearly
higher growth rate in the S than in the A solution.  

\section{Time--latitude diagram for minimal set of coefficients}
\label{butmin}

\begin{figure*}[h!]
\begin{center}
\includegraphics[width=0.7\textwidth]{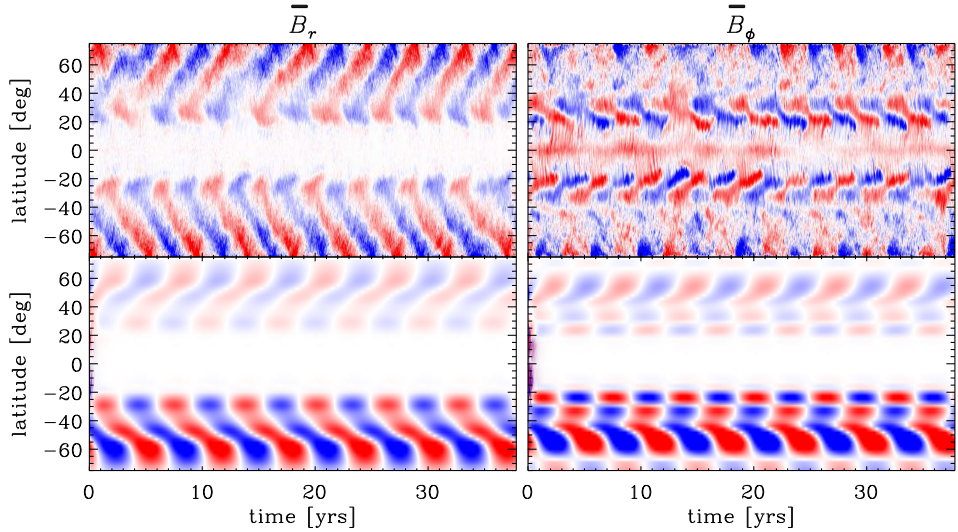}
\end{center}\caption{
Time--latitude (butterfly) diagrams of mean radial, $\meanB_r$, and
azimuthal, $\meanB_\phi$, magnetic field from the \MR simulation (top) and the MF
model (bottom) at fractional radius 0.95 for the minimal set of coefficients (cf. \FFig{but}).
The TTCs are symmetrized,
and $\aalpha$ is scaled by 1.5, 
while exponential growth has been compensated for clarity.
The color range is cut to make the 
northern hemisphere more visible.
}\label{butS2}
\end{figure*}

To show that the minimal set of coefficients,
$U_{\phi}$, $\aalpha$, $\gamma_{r,\theta}$, $\bbeta$, $\delta_{r,\phi}$, $\kappa_{\theta\theta r,\phi\theta r}$, 
is able to reproduce
the main features of and the dynamo mode of
the \MR simulation, we show in \FFig{butS2}  
for comparison time--latitude diagrams of the \MR simulation and the MF model.
Similar to the MF model including all coefficients, we find a very good agreement, most pronounced at midlatitudes.

\section{Solutions close to cycle period of \MR simulation}
\label{TabC}

In Table~\ref{ncyl} we show growth rates and oscillation frequencies of
those solutions that are close in cycle period to the \MR simulation with $\PGCD=4.4\pm0.6$ yr
(Class C). 
The solutions are labeled by their TTC selection.
We also define a quality factor $Q$ based on the cycle
period defined as 
\begin{equation}
Q=1-\left({P_{\rm cyc} -\PGCD}\over\PGCD\right)^2,
\label{qual}
\end{equation}
where the exact match of $P_{\rm cyc}$ and $\PGCD$ yields unity.
If the growth rate is negative (no dynamo) we set $Q$ to 0.

\begin{table}[h!]\caption{
Solutions with cycles between 3 and 5.8 yr, corresponding to $Q \ge 0.9$.
Symmetry indicates symmetric (S) and antisymmetric (A) solutions.
$\lambda$ is the growth rate of the volume-integrated rms
$\meanBB$, $P_{\rm cyc}$ is the period of the dominant oscillatory mode
and $Q$ is the quality factor defined in \Eq{qual}.
For each set of TTC selections (separated by solid horizontal lines)
all other tensors are fully active.
In the penultimate block all active TTCs are listed 
explicitly for the minimal set of coefficients 
while the last line shows the solution parameters of
the \MR simulation.
}\vspace{12pt}\centerline{
\begin{tabular}{l|ccll}
TTC Selection
& Symmetry & $\lambda$ (1/yr) &$P_{\rm cyc}$ (yr) & $Q$\\
\hline
\hline
\multirow{1}{*}{$\mean{U}_{\phi}$} & S/A & 0.85 & 4.86 & 0.99\\ 
\multirow{1}{*}{$\mean{U}_{\theta}$ $\mean{U}_{\phi}$} & S/A & 0.99 & 5.28& 0.96\\
\multirow{1}{*}{$\mean{U}_{r}$ $\mean{U}_{\phi}$} & S/A & 0.82 & 4.29& 1.00\\
\multirow{1}{*}{$\mean{U}_{r}$ $\mean{U}_{\theta}$ $\mean{U}_{\phi}$} & S/A & 0.92 & 4.54&1.00\\
\hline
\multirow{2}{*}{$\alpha_{rr}$ $\alpha_{r\theta}$ $\alpha_{r\phi}$ $\alpha_{\theta\theta}$ $\alpha_{\phi\phi}$} & S & 1.29 & 4.67&1.00\\
                        & A & 1.29 & 4.60&1.00\\[-3mm]
\multicolumn{5}{@{\hspace{-1mm}}c@{\hspace{-2mm}}}{\hdashrule{110mm}{.5pt}{.7pt 2pt}}\\[-1mm]
\multirow{1}{*}{$\alpha_{rr}$ $\alpha_{r\theta}$ $\alpha_{r\phi}$ $\alpha_{\theta\theta}$ $\alpha_{\theta\phi}$ $\alpha_{\phi\phi}$} & S/A & 0.92 & 4.54&1.00\\
$\alpha_{r\theta}$ $\alpha_{r\phi}$ $\alpha_{\theta\theta}$ $\alpha_{\phi\phi}$ & S & -0.12 & 5.69&0.00\\
\hline
\multirow{2}{*}{$\gamma_{r}$ $\gamma_{\theta}$} & S & 1.16 & 5.55&0.93\\
                    & A & 1.15 & 5.54&0.93\\[-3mm]
\multicolumn{5}{@{\hspace{-1mm}}c@{\hspace{-2mm}}}{\hdashrule{110mm}{.5pt}{.7pt 2pt}}\\[-1mm]
\multirow{1}{*}{$\gamma_{r}$ $\gamma_{\theta}$ $\gamma_{\phi}$} & S/A & 0.92 & 4.54&1.00\\
$\gamma_{r}$ & A & 1.30 & 3.25&0.93\\ 
\hline
\multirow{2}{*}{$\beta_{rr}$ $\beta_{r\phi}$ $\beta_{\theta\theta}$ $\beta_{\theta\phi}$ $\beta_{\phi\phi}$} & S & 0.56 & 4.00&0.99\\
                       & A & 1.13 & 4.01&0.99\\[-3mm]
\multicolumn{5}{@{\hspace{-1mm}}c@{\hspace{-2mm}}}{\hdashrule{110mm}{.5pt}{.7pt 2pt}}\\[-1mm]
\multirow{1}{*}{$\beta_{rr}$ $\beta_{r\theta}$ $\beta_{r\phi}$ $\beta_{\theta\theta}$ $\beta_{\theta\phi}$ $\beta_{\phi\phi}$} & S/A & 0.92 & 4.54&1.00\\
$\beta_{rr}$ $\beta_{\theta\theta}$ $\beta_{\phi\phi}$ & S & 0.34 & 4.00&0.99\\
$\beta_{rr}$ $\beta_{r\phi}$ $\beta_{\theta\theta}$ $\beta_{\phi\phi}$ & S & 0.37 & 3.78&0.98\\
$\beta_{rr}$ $\beta_{r\theta}$ $\beta_{r\phi}$ $\beta_{\theta\theta}$ $\beta_{\phi\phi}$ & S & 0.16 & 4.89&0.99\\
\hline
\multirow{1}{*}{$\delta_{r}$ $\delta_{\phi}$} & S/A & 0.94 &  4.86&0.99\\ 
\multirow{1}{*}{$\delta_{r}$ $\delta_{\theta}$ $\delta_{\phi}$} & S/A & 0.92 & 4.54&1.00\\
$\delta_{\theta}$ $\delta_{\phi}$ & A & 0.50 & 5.31&0.96\\
\hline
\hline 
\multirow{2}{*}{$\mean{U}_{\phi}$ $\aalpha$ $\gamma_{r}$ $\gamma_{\theta}$ $\bbeta$ $\delta_{r}$ $\delta_{\phi}$ $\kappa_{\theta\theta r}$ $\kappa_{\phi\theta r}$}& S  & 1.38 &  5.58&0.93\\ 
                   & A & 1.38 &  5.56&0.93\\ 
\hline
\hline
\MR & mixed & 0.00 & 4.4$\pm$0.6&1 to 0.98\\
\hline
\end{tabular}\label{ncyl}}
\end{table}

\end{document}